# 6RLR-ABC: 6LOWPAN ROUTING PROTOCOL WITH LOCAL REPAIR USING BIO INSPIRED ARTIFICIAL BEE COLONY

Nurul Halimatul Asmak Ismail[1], Samer A. B. Awwad[2], Rosilah Hassan[3]

[1]Department of Computer Science and Information Technology,
College of Community, Princess Nourah Bint Abdulrahman University
Kingdom of Saudi Arabia

[2]Independent Researcher, Kingdom of Saudi Arabia

[3]Center for Cyber Security,Faculty of Information Science and Technology,
UniversitiKebangsaan Malaysia, 43600, UKM Bangi, Selangor Darul Ehsan, Malaysia

## *ABSTRACT*

*In recent years, Micro-Electro-Mechanical System (MEMS) has successfully enabled the development of IPv6 over Low power Wireless Personal Area Network (6LoWPAN). This network is equipped with low-cost, low-power, lightweight and varied functions devices. These devices are capable of amassing, storing, processing environmental information and conversing with neighbouring sensors. These requisites pose a new and interesting challenge for the development of IEEE 802.15.4 together with routing protocol. In this work, 6LoWPAN Routing Protocol with Local Repair Using Bio Inspired Artificial Bee Colony (6RLR-ABC) has been introduced. This protocol supports connection establishment between nodes in an energy-efficient manner while maintaining high packet delivery ratio and throughput and minimizing average end-to-end delay. This protocol has been evaluated based on increasing generated traffic. The performance of the designed 6RLR-ABC routing protocol has been evaluated compared to 6LoWPAN Ad-hoc On-Demand Distance Vector (LOAD) routing protocol. LOAD protocol has been chosen since it is the most relevant existed 6LoWPANrouting protocol. The simulation results show that the introduced 6RLR-ABC protocol achieves lower packet average end-to-end delay and lower energy consumption compared to LOAD protocol.Additionally,the packet delivery ratio of the designed protocol is much higher than LOAD protocol. The proposed 6RLR-ABC achieved about 39% higher packet delivery ratio and about 54.8% higher throughput while simultaneously offering lower average end-to-end delay and lower average energy consumption than LOAD protocol.*

## 1. INTRODUCTION

Recent advances, in wireless sensor networks, are widely proliferated of embedded applications. The diversity of applications are ranging from smart mobility and smart tourism, public safety and environmental monitoring, smart home, smart grid, industrial processing, agriculture and breeding, logistics and product lifetime management, medical and healthcare, and independent living [1-2]. The emergence of a new paradigm, Low power Personal Area Networks (LoWPANs) is described well by the wireless standard IEEE 802.15.4. Internet Engineering Task Force (IETF) had managed an effort to integrate the standard of IEEE 802.15.4 networks and Internet Protocol version 6 (IPv6) called 6LoWPAN [3]. The rising of the new device embedded with internet connectivity is a dominant candidate for innovative networks [4].

6LoWPAN is principally challenging because of two technologies; the Internet Protocol (IP) and limited device standard IEEE 802.15.4 that collaborate together. The 6LoWPAN wireless devices are designed to have battery powered and need to keep low operation cycles [5]. Besides that, the encapsulated IPv6 packet, within the limited bandwidth and frame size, requires fragmentation and reassembly of data packets [6]. Hence, some algorithms had been proposed for header compression [7-8]. Routing protocol is used to ensure that the packet is sent from the source and received by the destination through the optimal paths. The optimal paths are measured in terms of some criteria such as number of hops, traffic, energy usage, bandwidth and shortest delay and able to work with limited power of nodes and limited capacity of the wireless link [9]. According to network structure, LoWPAN routing protocols had been classified into three main categories: flat-based, hierarchical-based and location-based routing protocols. On the other hand, routing protocols had been classified based on protocol operation into two types: Firstly, distance vector routing and secondly, link-state routing [5-6]. Additionally, routing protocols had been classified based on the updating style of routing tables of nodes to proactive, reactive and hybrid [10].

In 6LoWPAN, two routing schemes had been introduced: mesh-under and route-over. In mesh-under, the routing decision is not taken in the network layer, and hence it is not performed based on IP address. Conversely, the routing decision in mesh-under routing is performed at adaptation layer. Hence, it is performed based on the IEEE 802.15.4 Medium Access Control (MAC) addresses (16-bits or 64-bits).In contrast, in route-over, the routing decision between IPv6 domain and PAN is performed based on the IP address at the network layer [11].

In this paper, a new bio inspired swarm intelligence routing scheme has been proposed. This scheme is used to establish local route repair in order to overcome the link break between the source and the destination in 6LoWPAN.The scheme has its exclusive invent of a reliable routing protocol that considers the artificial bee colony algorithm of foraging behaviour and the link breakages. Simulation has been performed to evaluate the performance of our proposed scheme in comparison to 6LoWPAN Ad-hoc On-Demand Distance Vector (LOAD) routing. The rest of this paper is structured as follows: related works that summarize the LOAD, MLOAD and Originator Recognition (OR) path recovery mechanism for LOAD-Based Routing Protocol is presented in Section 2, background of 6LoWPAN routing protocols and their classification are introduced in Section 3,artificial bee colony algorithm is presented in Section 4, the introduced 6LoWPAN routing protocol with local repair using bio inspired artificial bee colony is discussed in Section 5, results are analysedin Section 6. Finally, the paper is concluded in Section 7.

## 2. RELATED WORKS

Nowadays, a number of approaches have been introduced in routing protocols to repair a link break. The approaches are depending on how the link is set up in the network and the reason of its failure [12].In 6LoWPAN, the changing of network topology causes links breaks. Hence, the active communication between the source and the destination is disconnected and fails to deliver the data. There are few 6LoWPAN flat-based routing protocols that try to repair the links breaks .LOAD, 6LoWPAN Ad Hoc On-Demand Distance Vector with Multi-Path Scheme (MLOAD) and Originator Recognition (OR) Path Recovery Mechanism for LOAD-Based Routing Protocol are the routing protocols that have mechanisms to repair links breaks.

LOAD tries to repair the link break locally by sending RREQ on behalf of the originator node. The intermediate repairing node that discovers the route break initiates a RREQ message on behalf of the originator to repair the broken link. It sets the originator address field of the RREQ message to its own address while maintains the destination address as in the original message. The intermediate repairing node sends the RREQ and waits until the destination responds the RREP. When the intermediate repairing node receives the RREP, the process of link local repair is completed. Hence, the buffered data packets are transmitted to the destination and routing table entry information is updated [13].However, when the intermediate repairing node does not receive a RREP, the intermediate repairing node unicasts route error (RERR) message with an error code specifying the cause of the repair failure back to the originator. Note that, RERR message must be within route error rate limit per second or the data packets are discarded. LOAD drawback is that it increases energy consumption as well as delay as the originator source needs to initiate a new route discovery process if the local repair does not work.

MLOAD broadcasts RREQ over the 6LoWPAN network to get to the destination. However, MLOAD adapts the concept of multi-path or multi-way to the same destination. Instead of having a single path to the destination, MLOAD needs three paths to reach the destination. Hence, when the primary route fails or the device fails, then the alternate route is used to transmit the data [14]. The limitation of MLOAD is that it increases the memory usage as the algorithm has to generate multiple paths to reach the same destination.

On the other hand, the Originator Recognition (OR) Path Recovery Mechanism for LOAD-Based Routing Protocol has three states to repair the link break between the source and the destination; memorizing, encapsulation and recognition. Each state triggers another state to be initialized. The RERR message with the added originator address header is used to identify which node will receive the RERR message. If the originator receives the RERR message, it reinitiates a new route discovery process for new path to continue transmitting the rest of the packets. However, if the receiving node of RERR message is not the source of the failed data packet, the RERR header field is discarded. Then, the RERR is forwarded to upstream node until it reaches the originator of the data packets [15]. This work can increase network overhead by having three states. Another routing protocol called Lightweight On-demand Ad hoc Distance-vector Routing Protocol-Next Generation (LOADng) has been developed for low power and lossy network. In this protocol, the route discovery process begins only after there is data needed to be received at the destination. Hence, this protocol minimizes both the routing overhead and the memory consumption [16].

Energy-Efficient Probabilistic Routing (EEPR) has developed the idea from AODV routing protocol. A RREQ packet in EEPR transmission has a specific forwarding probability that depends on the residual energy and the Expected Number of Transmissions (ETX) metric. Hence, the forwarding probability will be high if the ETX metric is low while the residual energy is high. In this work, the EEPR routing protocol shows that the variance of residual energy is lesser than the variance of the residual energy of AODV routing protocol [17].

The routing protocol highlighted above; LOAD, MLOAD and Originator Recognition (OR) Path Recovery Mechanism for LOAD-Based Routing Protocol have been designed specifically for 6LoWPAN, while the EEPR is designed for Internet of Things (IoT) environment. The difference between each routing protocol is the mechanism they use to repair the link break between the source and the destination. For example, LOADng uses the process of route discovery when there is in need of forwarding data to the destination while EEPR routing protocol did not consider to repair the link break. On the other hand, our workfocuses on bypassing the broken link.

# 3. BACKGROUND OF 6LOWPAN ROUTING PROTOCOL

The state of the art of 6LoWPAN routing protocols can be classified depending upon network structure or protocol operation. However, specific routing protocol can be hybrid from these categories.

## 3.1 Network structure routing

Based on the network structure, routing protocols can be classified into the following types: flat-based routing [18], hierarchical-based routing [19] and location based routing [20]. In this paper, the flat-based routing is adopted. In flat-based routing, 6LoWPAN nodes typically act and collaborate together to perform the task. The 6LoWPAN devices; full function device (FFD) and reduced function device (RFD) communicate with PAN coordinator which is also a FFD.

## 3.2. Route-over versus mesh-under routing

Based on the protocol operation, routing protocols can be classified into the following types:

- Mesh-under routing

Routing in mesh-under scheme requires a group of 6LoWPAN nodes and one edge router or so-called gateway that acts as an IPv6 router [21].The IPv6 address of each LoWPAN device is mapped to the IEEE 802.15.4 link-layer address. The routing decision in mesh-under routing is performed in the 6LoWPAN adaptation layer. The packet size could be huge, and hence the adaptation layer may need to fragment the packet. The adaptation layer prepares the 6LoWPAN encapsulated datagram and then passes the datagram to the MAC layer. The MAC layer encapsulates the received datagram within the IEEE 802.15.4 MAC frame. Then, the frame is sent through multiple link-layer hops to the destination [5][7][22]. In case the packet is fragmented, the fragments may go through different routes and they will be reassembled at the destination node. The missing fragments may cause the packet to be dropped or retransmitted [11]. In case the packet is originated from or destined to an external device, the edge router receives the IPv6 packet and then forwards the packet to its particular destination. LOAD and MLOAD are two popular examples of mesh-under routing protocols [23-24].

- Route-over routing

In route-over scheme, the routing decision is performed by the 6loWPAN network layer. In this routing scheme, every link-layer hop is an IP hop, and every node is an IP router to route the packet to the destination node. Hence, the encapsulated IPv6 packet is sent all the way through a multiple IP hops to the destination. Every IPv6 packet is reassembled and fragmented again at each hop. The adaptation layer reassembles the packet, passes the packet to the network layer for routing processing. After the network layer takes the routing decision, it passes the packet down again to the adaptation layer. The adaptation layer fragments the packet again and passes it to the MAC layer. However, a retransmission mechanism is in need to retransmit all or any missing fragments at the intermediate nodes [18][11][25].Recently, an enhanced route-over routing protocol called 6LoWPAN Route-Over with End-to-End Fragmentation and Reassembly Using Cross-Layer Adaptive Backoff Exponent has been developed. This protocol avoids the hop-by-hop fragmentation and reassembly to reduce the delay. In this protocol, the fragmentation is carried out at the source node only and fragments are reassembled at the destination node only. Additionally, the protocol uses cross layer design between adaptation layer and MAC layer to control the back off exponent based on the number of fragments in the packet. This reduces the fragments losses, and hence increases the packet delivery ratio [26].

## 4. ARTIFICIAL BEE COLONY ALGORITHM

Several studies had been introduced to utilize the Artificial Bee Colony (ABC) for wireless sensor network (WSN). However, ABC is yet to be utilized in 6LoWPAN network. The ABC is an optimization algorithm inspired by the intelligent foraging behaviour of honey bee [27]. Optimization problem is designed to find the best solution from all possible solutions. ABC algorithm has two points; first, the positions of food sources as the possible solutions and second, the nectar richness as the quality (fitness) of the associated solution. Additionally, the number of the employed bees or the onlooker bees is equal to the number of solutions in the population.

A research on the ABC algorithm is adopted and adapted for sensor deployment problem; sensor deployment problem in 3-D terrain, dynamic deployment problem for mobile sensor networks and routing operations in WSN [28-32].Other than that, BEES is a lightweight bio inspired backbone construction protocol, which can help mitigate sensor localization, clustering, and data aggregation among others [33]. This protocol had been proposed to simplify many of network management tasks like leader election, task management, and routing. BEE-C is a hierarchical routing protocol that has adapted the behaviour of bees in their work. It had been proposed to save the energy of sensor nodes in WSN [34].

Bee Inspired Routing Protocol Using Lossless Compression Based on Swarm Technology had been introduced [35]. The protocol uses the concept of swarm intelligence of bee-inspired dependent to support multipath routing wireless ad hoc network. It is emphasized in order to find a number of routes and spread the traffic in a proper way. However, to the best of our knowledge, there is no research has utilized the adopting of ABC technique into 6LoWPAN routing.

## 5. 6LOWPAN ROUTING PROTOCOL WITH LOCAL REPAIR USING BIO INSPIRED ARTIFICIAL BEE COLONY

Link break is one of the main challenges that the routing protocols experience [13-15][36]. It suspends the traffic flow or totally stops it sometimes. Hence, routing protocol design should consider the link break to either find an alternative route or locally handle the problem by providing link local repair that replaces only the broken part of the route. As mentioned earlier, LOAD protocol is suitable for a dynamic self-starting and loop free route in mesh networks. This special algorithm is designed to re-establish route through locally repairing the broken link that may occur between intermediate forwarding nodes during actual data transmission. It maintains a routing table at each node and performs its task in two phases; route discovery and route maintenance phases [37]. However, LOAD has significant drawbacks in terms of delay and the overhead that spans for large area when performing the link local repair.

In this section, a new mesh-under routing protocol called 6LoWPAN Routing Protocol with Local Repair Using Bio Inspired Artificial Bee Colony (6RLR-ABC) is proposed. 6RLR-ABC has two phases; route discovery and route maintenance. The main contributions of this routing protocol are: first, reducing the size of link local repair area through bypassing the failure node only. Second, introducing a new bio inspired swarm intelligence local repair mechanism called LR-ABC for mesh-under routing protocol to select the best local link.

The 6RLR-ABC network architecture is shown in Figure 1. It has the following components: Source Node (SN): is the source of data packet. Destination Node (DN): is the intended destination of the data packet. Upstream node (UN): is the node that discovers the broken link. It is located at the beginning of the broken link. Abandoned Node (AN): is the intermediate failed node that causes the link break. Second Next Hop Node (SNH): is the intermediate node that is located after the abandoned node in the route. Intermediate Node (IN): is the node between the upstream and the second next hop nodes. Broken link: is the broken link between the upstream node and the abandoned node. Lastly, Link local repair: is the link that connects the upstream node and the second next hop node. It bypasses the abandoned node to avoid the broken link.

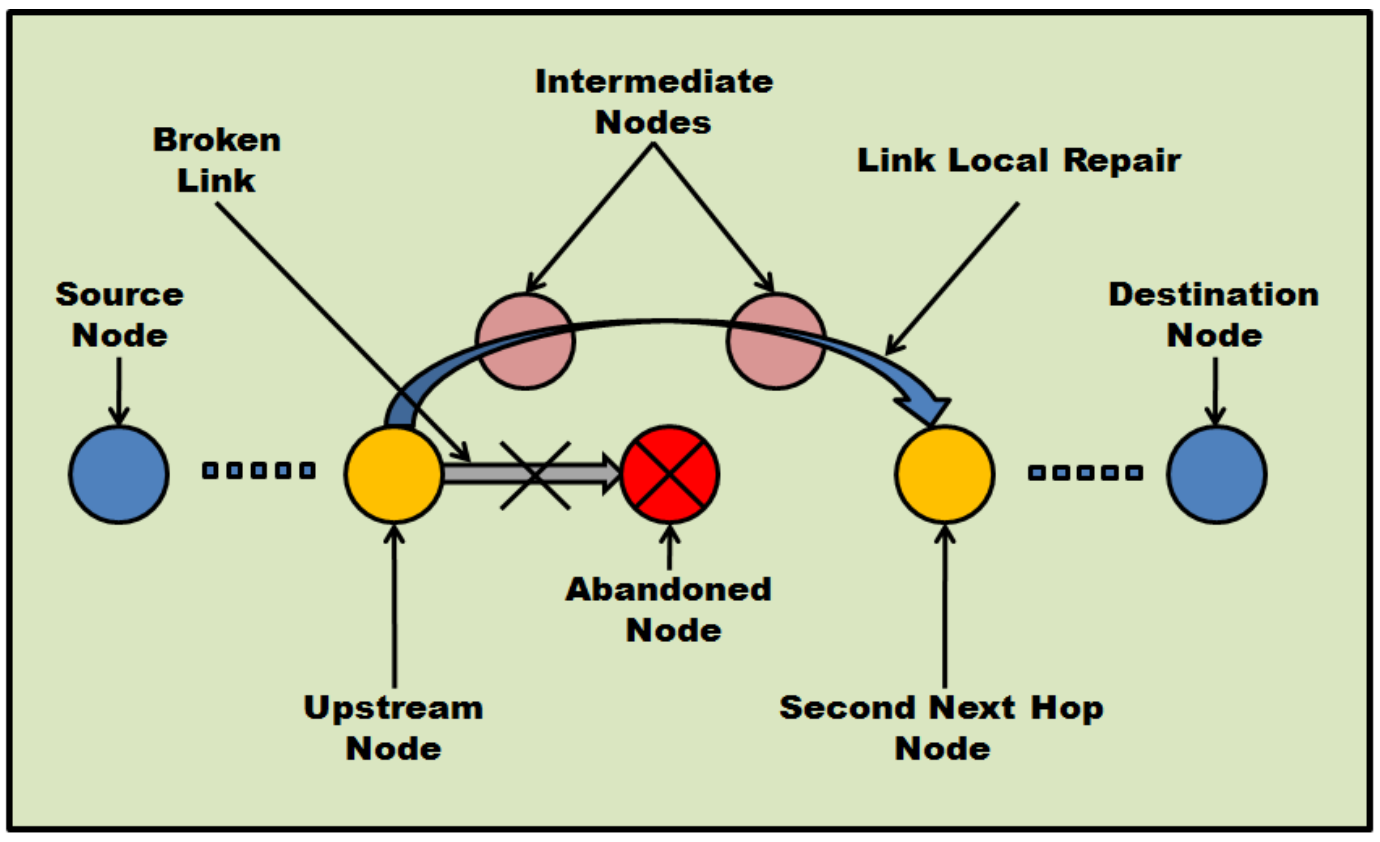

Figure 1. The 6RLR-ABC network architecture and components

In LR-ABC mechanism, the upstream node, that discovers the link break, initiates Local_RREQ message on behalf of the source node. It sends the Local_RREQ to the second next hop node located after the failure node. This limits the link local repair area and controls the overhead of the process. The link local repair areas for both LR-ABC mechanism in 6RLR-ABC and LOAD routing protocols are shown in Figure 2.

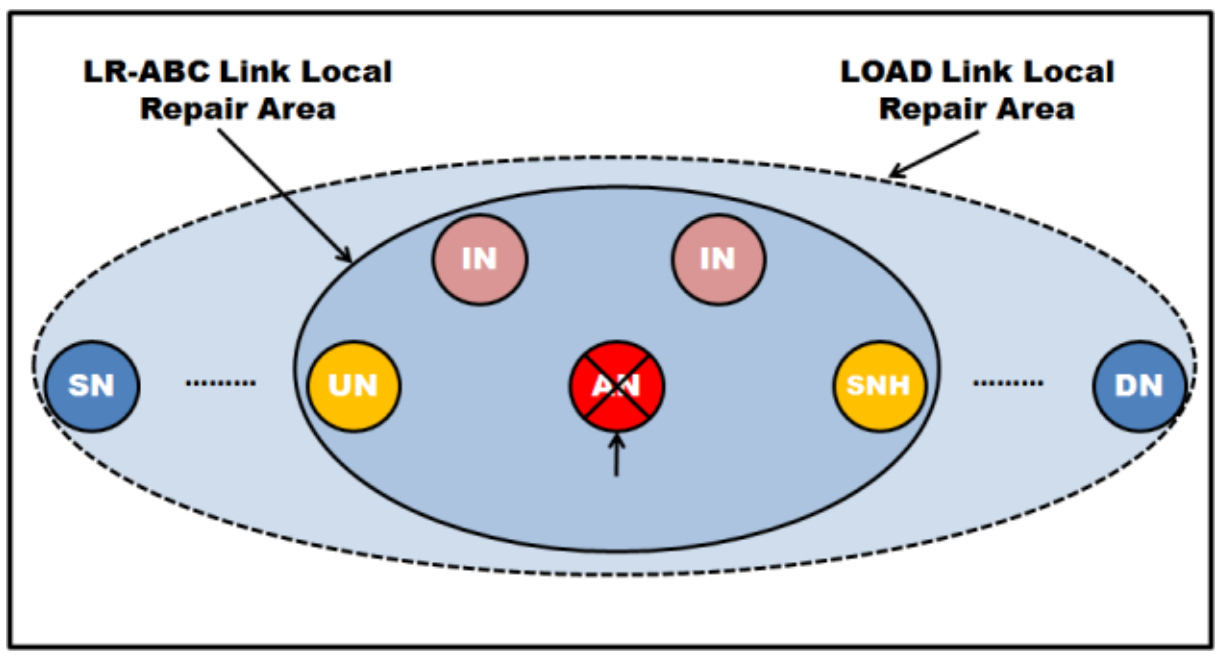

Figure 2. The link local repair areas for both LR-ABC mechanism in 6RLR-ABC and LOAD routing protocols

In addition, the LR-ABC mechanism selects the best local link repair based on the amount of the residual energy for the entire nodes in the local link.The best local link selection in LR-ABC mechanism is shown in Figure 3. The 6RLR-ABC mesh-under routing protocol and LR-ABC mechanism are discussed in the following subsections. The packet's format is presented first. After that, route discovery is introduced. Finally, the LR-ABC mechanism is explained.

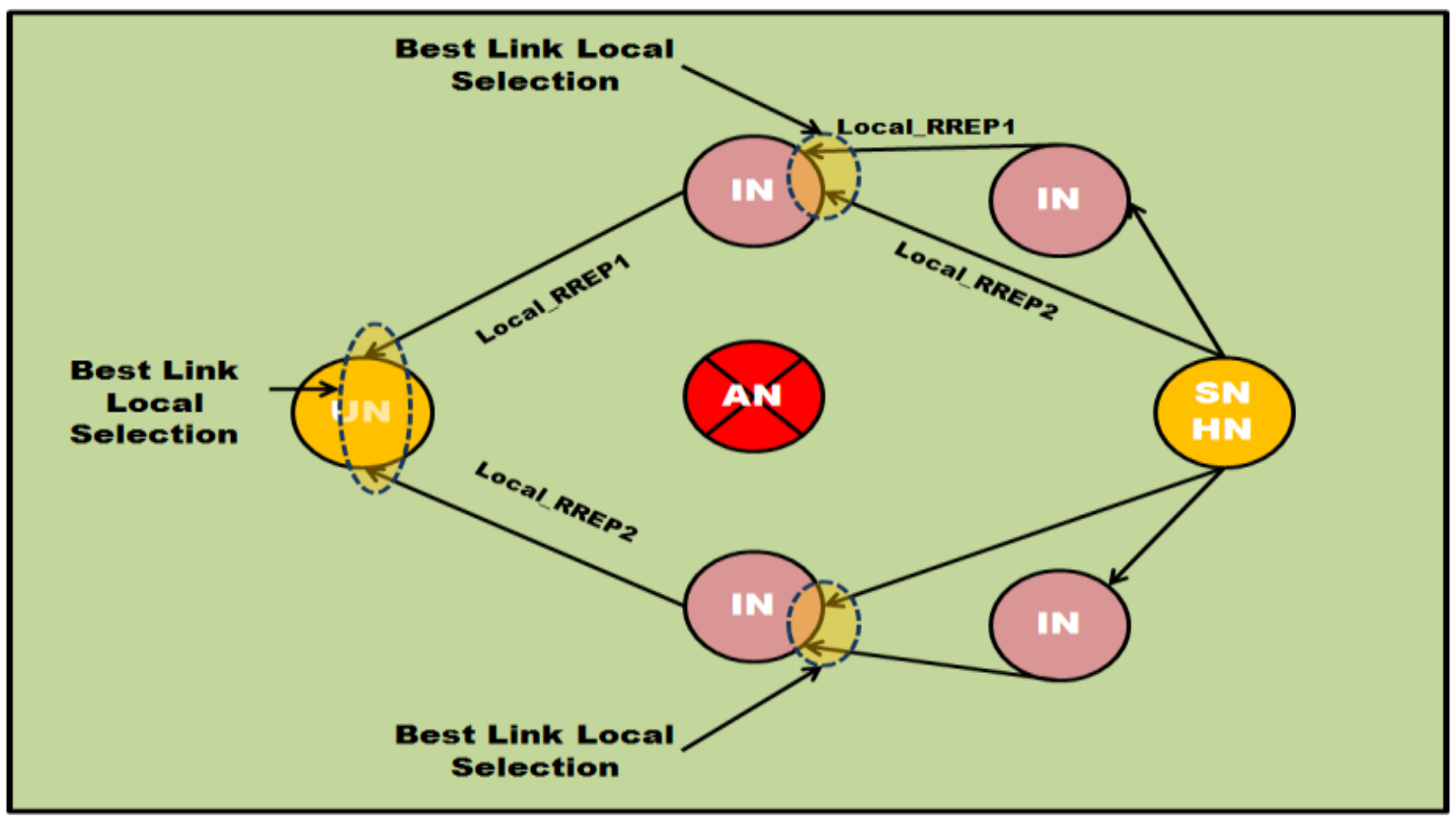

Figure 3. The Best Local Link Selection in LR-ABC mechanism

## 5.1 Packet Format

6RLR-ABC is designed based on bio inspired ABC. Hence, the packet format in 6RLR-ABC routing protocol comprises the information that is needed by the bio inspired ABC mechanism. In this packet format, LOAD packet format is modified by adding Hop Count, Accumulated residual Energy Level (AEL), RREP_ID and Link Layer Second Next Hop Address in both RREQ and RREP packet format. The 6RLR-ABC packet format is shown in Figure 4.

<table>
<tr><th colspan="2">8</th><th>1</th><th>1</th><th>1</th><th>5</th></tr>
<tr><td colspan="2">Type</td><td>R</td><td>D</td><td>O</td><td>Hop Count</td></tr>
<tr><td>CT</td><td>WL</td><td colspan="4">AEL</td></tr>
<tr><td colspan="2">RREQ_ID</td><td colspan="4">RREP_ID</td></tr>
<tr><td colspan="6">Link Layer Destination Address</td></tr>
<tr><td colspan="6">Link Layer Originator Address</td></tr>
<tr><td colspan="6">Link Layer Second Next Hop Address</td></tr>
</table>

Figure 4. RREQ and RREP packet format

The Hop Count field represents the number of hops from the source or the originating node to the current node that handles the message. This value is initially set to 0 and incremented by one at each intermediate node visited by the message. When the Hop Count value reaches a predefined limit, the message will be discarded. The AEL represents the accumulated residual energy level for all the nodes along the way between the RREQ or RREP message initiator and the current node that has received the message. The total number of Weak Links (WL) and type of Route Cost (CT) originally from LOAD. WL indicates the total number of weak links on the routing path from the originator to the sender of the RREQ in RREQ message or from the originator of the RREP to the sender of the RREP in RREP message. The CT indicates the type of route cost. TheRREQ_ID and RREP_ID are unique sequence numbers used to identify RREQ and RREP messages to the source and the destination nodes respectively. Lastly, Link Layer Second Next Hop Address is the link layer address of the next node after the failure node.

### 5.2 Route Discovery in 6RLR-ABC

When the source node needs to send data to the destination while the route to the destination is unavailable, the route discovery phase in 6RLR-ABC is started. The route could be unavailable at the source node because of two reasons; the destination is unknown to the source or the route to the destination had expired or became invalid. In the route discovery process, firstly, the originator broadcasts a RREQ message to the neighbouring nodes. Each node that receives the RREQ updates its routing table entry and forwards the RREQ to other nodes. If one of the intermediate nodes has a valid route to the destination, it directly replies to the source. Upon receiving the RREQ, the intermediate node that knows the route to the destination or the destination node itself sends back RREP message to the source through the route that has been established during the journey of the RREQ. The RREP message carries the needed information about the route. Then, the source node uses the discovered route to send data to the destination.

### 5.3 Route Maintenance in 6RLR-ABC

When link failure occurs, the data transmission is suspended. Hence, routing protocol design should handle this problem by either initiating new route discovery or performing route maintenance for the existed broken route. However, route maintenance gains more interest due to its ability in repairing the broken link locally with low overhead. In addition, the route maintenance is performed within a short time compared to initiating a new route discovery process. The more locally the repair the more efficient it is. Hence, many local repair and maintenance mechanisms try to limit the maintenance area in order to achieve better performance through avoiding both the high overhead and the high delay of the maintenance process.

This work aims to locally repair the link break through limiting and controlling the maintenance area to the neighbouring nodes by bypassing the broken node only. The work intends to avoid involving both the source and the destination nodes in the route repair process. The source and the destination nodes could be located far from the failure nodes. Hence, involving them in the repair process may span the maintenance process over large area.ABC has inspired this work since the bees use the same phenomena when the food source is finished or broken. After the food source is vanished, finished or broken, the bees try to locate a new food source within the local vicinity. The new located food source will be used to supply the colony with the food source. ABC has four phases; initialization, scout bee, employed bee and onlooker bee phases. These phases and their implementation in LR-ABC mechanism are discussed as follows:

- Initialization phase

In the initialization phase, there are three ABC parameters that need to be initialized. These parameters are; the colony size (CS) is reciprocal of the number of nodes that are involved in the link local repair, food sources are reciprocal of the possible paths discovered in the link local repair process and the number of trials is reciprocal of the number of iterations in the local repair.

- Scout Bee phase

In this phase, the scout bees search for new food sources in the neighbouring area in order to replace the old vanished food source. In each scout bee's phase, each scout flies over one of the j potential areas to search for the new food source. In LR-ABC mechanism, scout bee is the upstream node or any intermediate node that disseminates or broadcasts Local_ RREQs messages during its search for the local route. As the scout bees search for new food sources in the neighbouring areas, the upstream and the intermediate nodes search for new local links to replace the broken link that contains the abandoned node. The upstream node broadcasts

Local_RREQ message toward all j neighbouring nodes to search for second next hop node. The upstream and the intermediate nodes enter the scout bee phase when they broadcast Local_RREQs messages. The nodes are remaining in this phase until they receive a Local_RREP message from the second next hop node.

After the intermediate node receives Local_RREQ, it discards the message if the message with the same source address and RREQ_ID was received before. Otherwise, it changes its status to scout bee phase. It then calculates its current Energy Level (EL) which indicates the residual energy level of the node and the AEL which indicates the average accumulative battery residual power for all nodes along the potential path to the source. After that, it compares the calculated AEL for the new link to the source with the existed one (if available) and the best local link is selected .It then shares its food source information with other nodes; it prepares the Local_ RREQ with the new calculated AEL, WL and CT fields, updates the second next hop address and broadcasts it to other neighbouring nodes. After the neighbouring node receives the Local_ RREQ message, it repeats the same process and broadcasts the Local_ RREQ to the other nodes. The process continues until the Local _RREQ is received by the second next hop node.

When the Local _RREQ is received by the second next hop node, a food source is considered found. The food source, in LR-ABC mechanism, is considered the path to get to the second next hop node (not the second next hop node itself).This local route is called LRj. The second next hop node responds to the received Loc al_RREQ by unicasting a Local _RREP message back toward the upstream node. When the upstream or intermediate node receives the Local_ RREP, it changes its status from scout bee to employed bee phase. Figure 5 shows the beginning and the ending of scout bee phase for the upstream node.

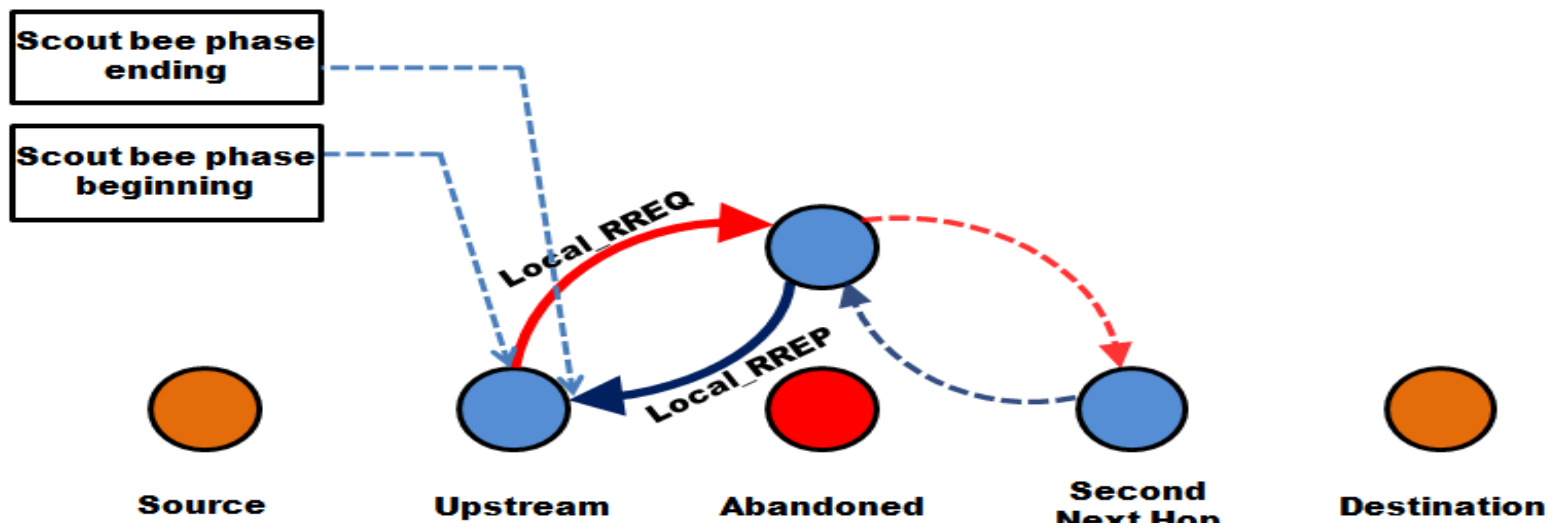

Figure 5. The beginning and the ending of scout bee's phase for the upstream node.

- Employed Bee phase

As mentioned in previous section, after the scout bee finds a new food source, its status is changed to the employed bee's phase. Then, it calculates the fitness of the new food source. After that, it compares the fitness of the new food source with the existed one (if available) and the best one is selected. Then, the employed bee goes back to the hive and shares its information about the food source with the onlooker bees.

The scout bee is changed to employed bee after it finds a new food source. Similarly, the upstream and the intermediate nodes enter the employed bee phase after they receive the Local_RREP message which is initiated by the second next hop node as a response to the Local_RREQ broadcasted earlier in the search process. The nodes remain at this phase until they receive new Local_RREQ message to start a new search for a food source (new local link repair process).

After the intermediate node receives Local_RREP, it changes its status to employed bee's phase. It calculates its current EL which indicates the residual energy level of the node and the AEL which indicates the average accumulative residual battery power for all nodes along the potential path to the second next hop node.It then, compares the calculated AEL for the new local link with the existed one (if available) and the best local link is selected. After that, it shares its food source information with previous node in the precursor list; It prepares the Local_RREP with the new calculated AEL, WL and CT fields, updates the second next hop address and shares (passes) it to the next node in the precursor list (Onlooker node). After the onlooker node receives the Local_RREP message, it repeats the same process and passes the Local_RREP to the next node in the precursor list (next onlooker bee). The process continues until the Local_RREP is received by the upstream node (anchor onlooker bee). Figure 6 shows the beginning and the ending of employed bee's phase for the intermediate node.

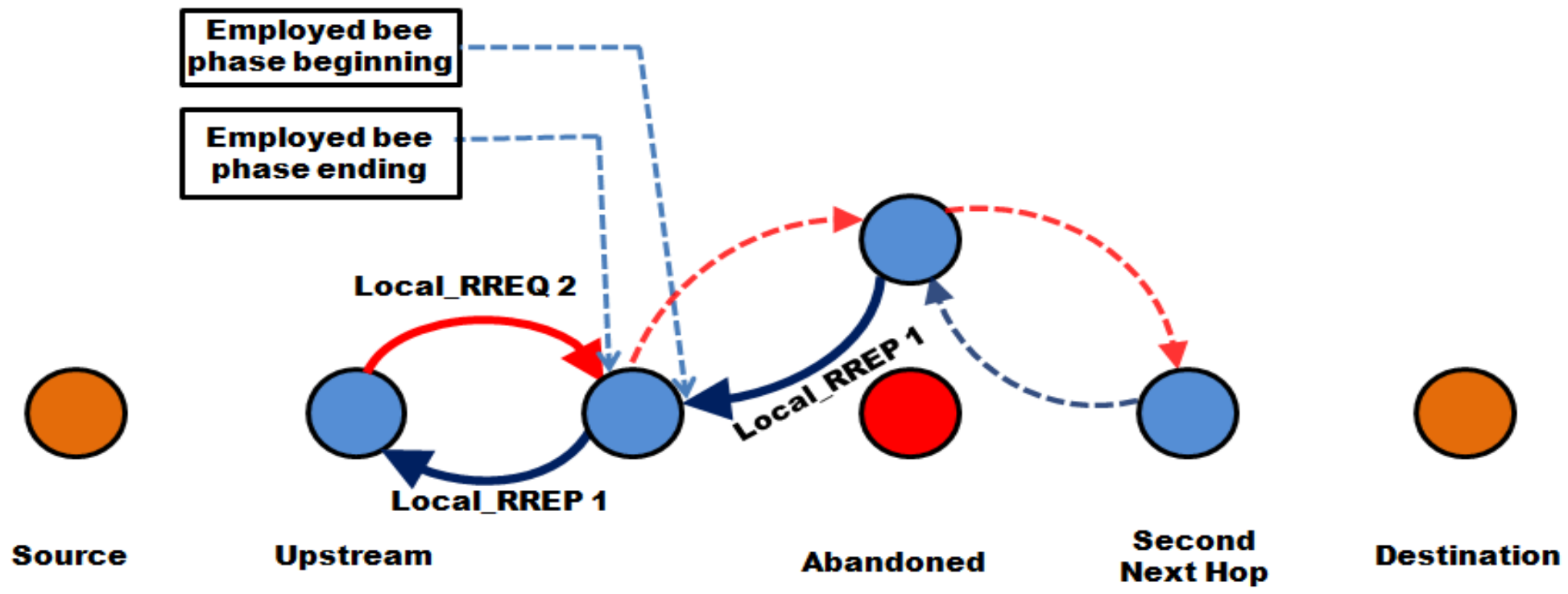

Figure 6. The beginning and the ending of employed bee's phase for the intermediate node.

- Onlooker Bee phase

After the employed bees complete their searches and return to the hive, they share their information about the quality of their new food sources with the onlooker bee. The onlooker bee chooses the food source based on the nectar amount (fitness) of the food sources shared by employed bees. The onlooker node calculates the fitness of the newly shared food source. It compares the calculated fitness with the existed one and the best food source is selected.

In LR-ABC mechanism, after the second next hop node or the intermediate node prepares and updates the Local_RREP, it sharesthe Local_RREP with the previous node in the precursor list. This node is considered as onlooker node from previous node point of view. This node plays the role of both onlooker and employed bees. As explained earlier, the node performs the task of employed beeto find the local link after receiving the Local_RREP. In addition, it searches for the best local link established by different Local_RREPs received from different nodes. It calculates the AEL for the new local link and compares it with the existed one.If the new discovered local link is better than the existed one, the node shares its information about the new local link with the previous onlooker node in the precursor list; it forwards the Local_RREP to the next node in the precursor list.

Otherwise, the Local_RREP is discarded. The process is continued until the upstream node (last or anchor onlooker node) receives the Local_RREP. The upstream node calculates the AEL for the new local link. If the new discovered local link is better than the existed one, it updates the local link to the new one. When the local link searching timeout is expired, the available local link in the route request table is considered the best local link. After that, the data transmission can be resumed. Figure 7 shows the beginning and the ending of onlooker bee's phase for the intermediate node (the next node in the precursor list).

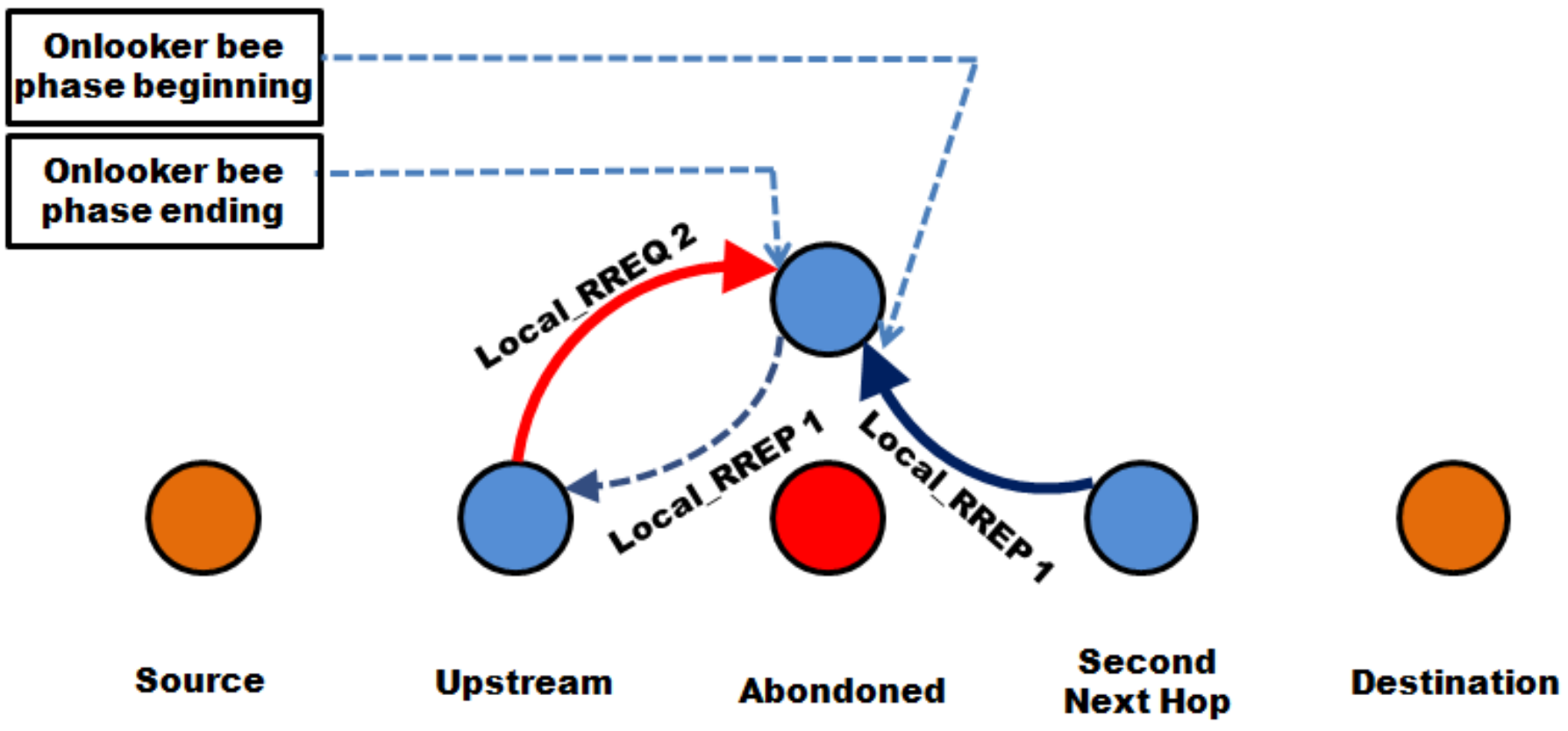

Figure 7. The beginning and the ending of onlooker bee's phase for the intermediate node.

## 6. Results and Discussions

In 6LoWPAN, the network spans through large geographical area. The source and destination use multi hops architecture to communicate with each other. Hence, various numbers of intermediate nodes may be presented between them. The main parameters that affect the performance of route-over and mesh-under routing protocols in 6LoWPAN are the number of hops between source and destination and the number of fragments in the IPv6 packet. Hence, the performance of 6LoWPAN routing protocols is examined, tested and evaluated based on multi hops chain and multi fragments scenarios [11][38].While, Chained MUR(C-MUR) 6LoWPAN routing protocol had been evaluated based on chain, linear, five hops scenario [39], the second and subsequent fragments header compression technique had been evaluated in multi hops chain and multi fragments scenarios [7-8].

For performance evaluation, the Qualnet simulator has been used to conduct the simulation experiments. The simulation results of the proposed 6RLR-ABC routing protocol have been compared to the simulation results of the LOAD routing protocol in 6LoWPAN network. Figure 8shows the scenario whereby the link failures occur nearby the source node. Table 1 shows the main simulation parameters that are configured during the simulation experiments. These simulation parameters are based on previous word presented earlier [40].

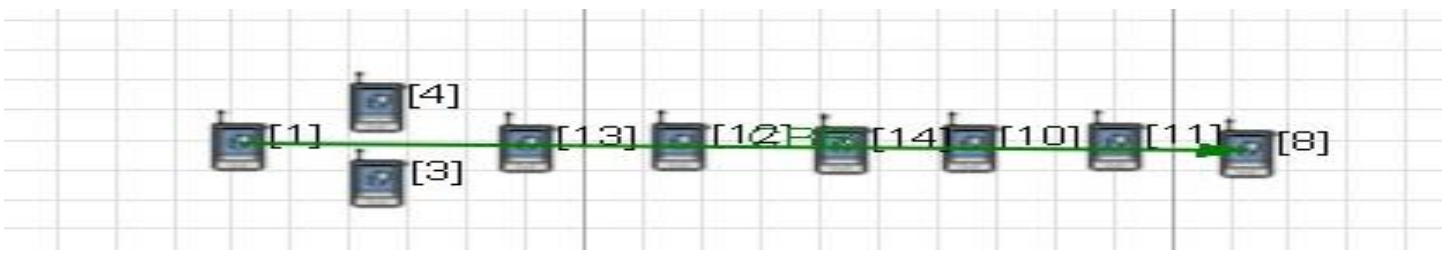

Figure 8. The simulation scenario whereby the link failures occur nearby the source node

Table 1. Simulation parameters used in the simulation

| Parameter | Value |
|---|---|
| Radio type | 802.15.4 radio |
| MAC protocol | 802.15.4 |
| Network protocol | IPv6 |
| Routing protocol | LOAD, 6RLR-ABC |
| No of nodes | 9 nodes |
| Scenario dimension | 1500 X 1500 m2 |
| Simulation time | 1020 sec |
| Packet size | 50 |
| Node placement model | Random |
| Application protocol | CBR |
| Number of simulations | 100 |

The following four metrics are considered for the performance evaluation: Average energy consumption, throughput, packet delivery ratio and average end-to-end delay.

## 6.1 Average Energy Consumption

The 6RLR-ABC has lower average energy consumption compared to LOAD. The reason is that the 6RLR-ABC has less overhead compared to LOAD which conserves the energy used to initiate both RREQ and RREP messages between the source and the destination during route establishment.

After the link break, the LOAD protocol locally repairs the broken link through sending a Local_RREQ message to the destination which is located far (7 hops) from the upstream node.The upstream node, in this scenario, is the source node itself. Hence, the Local_RREQ is broadcasted toward the destination which is 7 hops far from the destination. In addition, the Local_RREP message passes the same number of hops after responded by the destination. The consumed energy in local link repair in LOAD ($E_{LLR\text{-}LOAD}$) can be simply expressed in the following equation:

$$E_{LLR-LOAD} = ND_{LLR-LOAD} * (E_{LRREQ-t} + mE_{LRREQ-r}) + H_{U-D} * (E_{LRREP-t} + E_{LRREP-r})$$

Where: $ND_{LLR\text{-}LOAD}$ is the network diameter that specifies the number of nodes that transmit and receive the Local_RREQ. Since LOAD does not consider the Local_RREQ dissemination problem,this variable may include all nodes in the network.$E_{LRREQ\text{-}t}$ is the energy consumed in broadcasting a Local_RREQ message to the neighbours. $E_{LRREQ\text{-}r}$ is the energy consumed in receiving a Local_RREQ message. $m$ is the average number of neighbours that forward the Local_RREQ.$H_{U\text{-}D}$ is the number of hops between the upstream node and the destination node. $E_{LRREP\text{-}t}$ is the energy consumed in transmitting a Local_RREP message. $E_{LRREP\text{-}r}$ is the energy consumed in receiving a Local_RREP message.On the other hand, the consumed energy in link local repair in 6RLR-ABC ($E_{LLR\text{-}LABC}$) can be simply expressed in the following equation.

$$E_{LLR-LABC} = ND_{LLR-LABC} * (E_{LRREQ-t} + mE_{LRREQ-r}) + H_{U-S} * (E_{LRREP-t} + E_{LRREP-r})$$

Where: $ND_{LLR\text{-}LABC}$ is the network diameter that specifies the number of nodes that transmit and receive the Local_RREQ. Since the second next hop is few hops far from the upstream node, the number of nodes that broadcast the Local_RREQ is controlled. $H_{U\text{-}S}$ is the number of hops between the upstream node and the second next hop.

In the current scenario, as an example:

$$E_{LLR-LOAD} = 9 * (E_{LRREQ-t} + 2E_{LRREQ-r}) + 7 * (E_{LRREP-t} + E_{LRREP-r})$$

and

$$E_{LLR-LRABC} = 4 * (E_{LRREQ-t} + 2E_{LRREQ-r}) + 2 * (E_{LRREP-t} + E_{LRREP-r})$$

As can be shown from these two equations, 6RLR-ABC conserves the energy used in local link repair. It uses (4/9≈0.44) in broadcasting the Local_RREQ message compared to LOAD. In addition, it uses (2/7≈0.29) during the Local_RREP response compared to LOAD.

Table 2 and Figure 9 show the average energy consumption versus the number of transmitted packets for both LOAD and 6RLR-ABC protocols. The figure depicts that; 6RLR-ABC outperforms LOAD in terms of average energy consumption. As the figure shows, in overall average, 6RLR-ABC has up to 17% lower average energy consumption compared to LOAD.

Table 2. The Average Energy Consumption versus Traffic

| Traffic | 6RLR-ABC Average Energy Consumption (mAhr) | LOAD Average Energy Consumption (mAhr) |
|---|---|---|
| 100 | 0 | 0 |
| 200 | 0.1592224 | 0.1926591 |
| 300 | 0.0829993 | 0.0987691 |
| 400 | 0.053332 | 0.0645318 |
| 500 | 0.0373604 | 0.0440853 |
| 600 | 0.0299639 | 0.0356571 |
| 700 | 0.0250881 | 0.0301057 |
| 800 | 0.0220506 | 0.0262402 |
| 900 | 0.01957 | 0.0232883 |
| 1000 | 0.0175531 | 0.0212392 |
| 1100 | 0.016211 | 0.0196153 |
| 1200 | 0.0150369 | 0.0177436 |

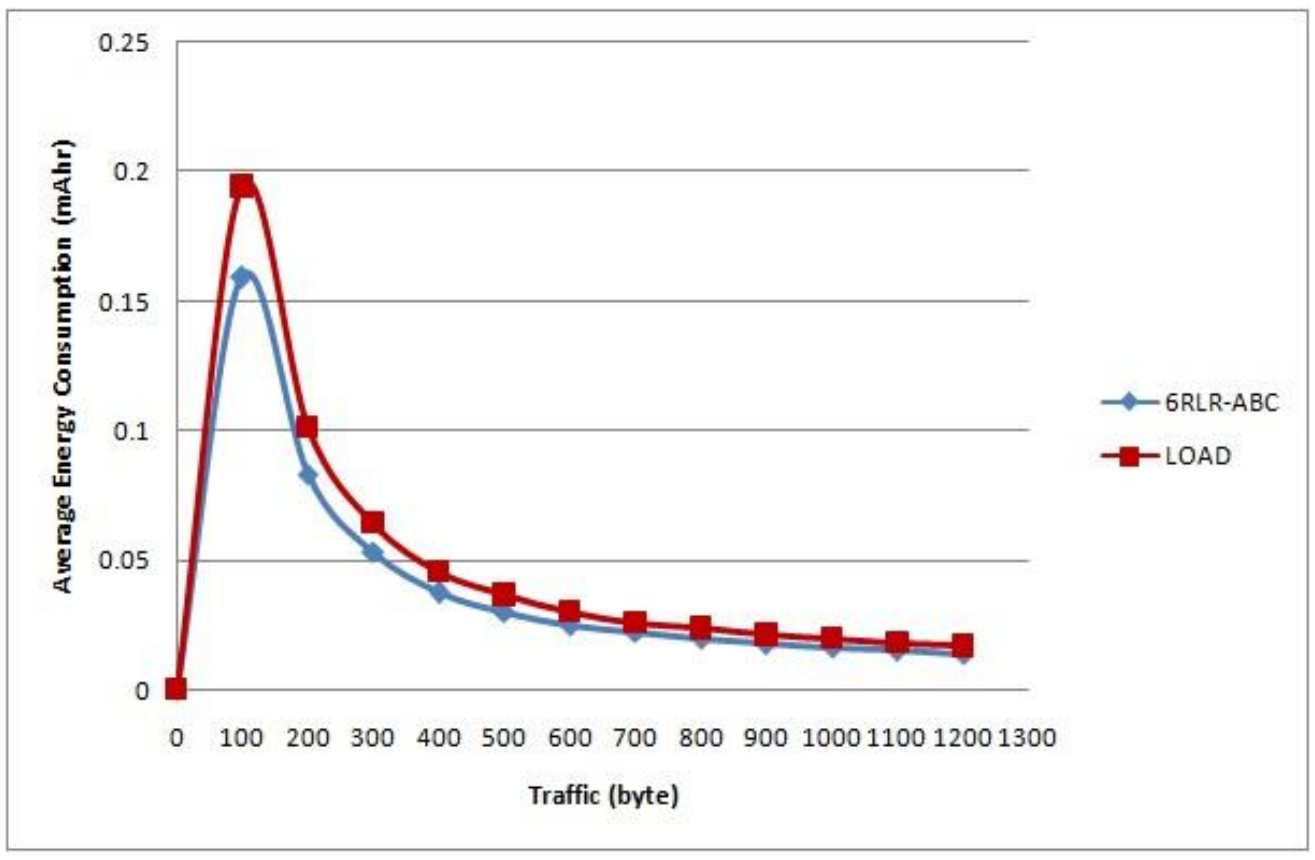

Figure 9.Average energy consumption versus traffic

## 6.2 Throughput

Table 3 and Figure 10 show the throughput for both protocols. As can be noted from the figure, when the traffic increases, the number of collisions increases as well. This explains the reduction of the throughput for both protocols. Nevertheless, 6RLR-ABC has higher throughput compared to LOAD. It sends lower overhead compared to LOAD, and hence it has lower collisions and packets losses.6RLR-ABC achieves more than 54.8% higher throughput compared to LOAD.

Table 3. Throughput versus Traffic

| Traffic | 6RLR-ABCThroughput(bps) | LOADThroughput(bps) |
|---|---|---|
| 100 | 0 | 0 |
| 200 | 51.533970 | 41.333970 |
| 300 | 99.728890 | 79.228890 |
| 400 | 151.780900 | 119.080900 |
| 500 | 217.627600 | 168.727600 |
| 600 | 271.448400 | 207.348400 |
| 700 | 320.619500 | 242.219500 |
| 800 | 366.081700 | 272.781700 |
| 900 | 413.668500 | 304.368500 |
| 1000 | 458.800700 | 333.000700 |
| 1100 | 491.741900 | 336.741900 |
| 1200 | 536.720200 | 355.920200 |

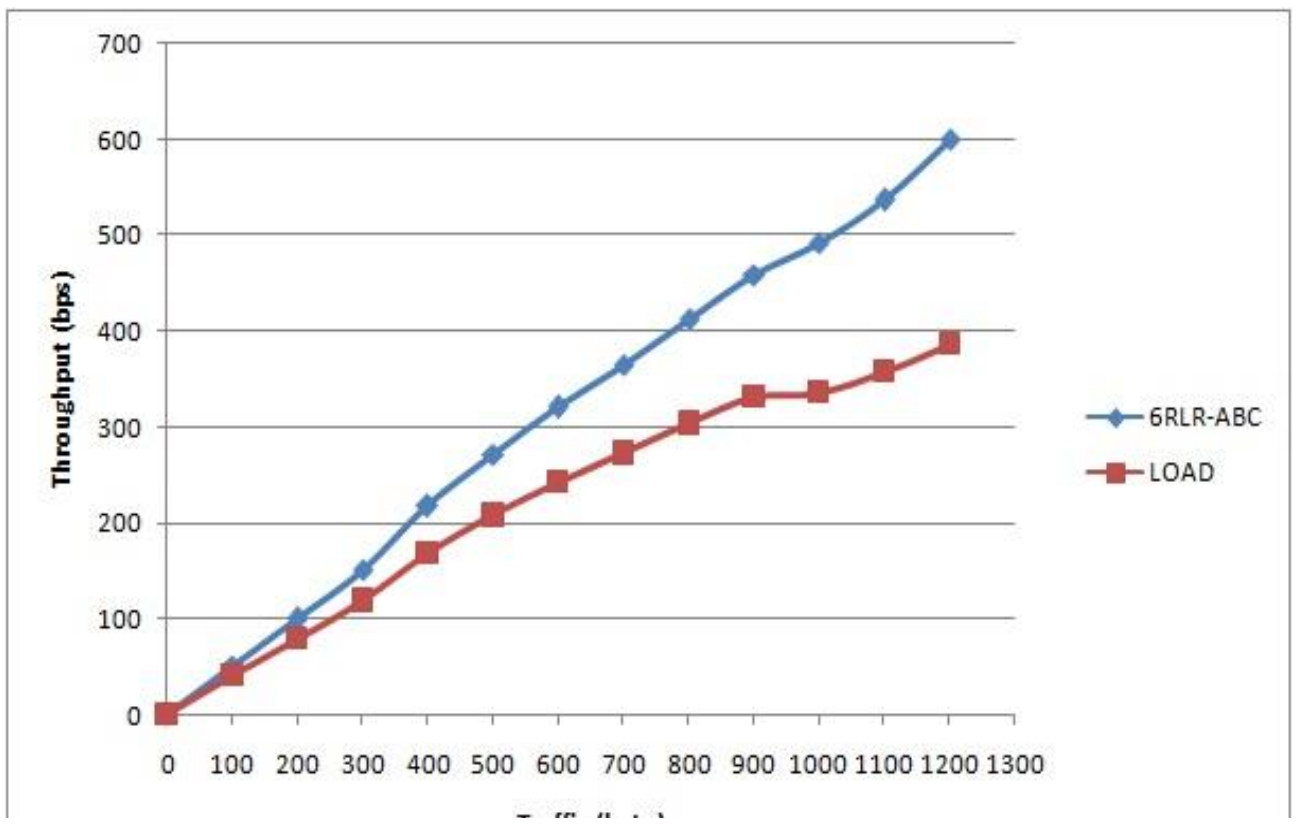

Figure 1. Throughput versus Traffic

## 6.3 Packet Delivery Ratio

Once link failure occurs, the LOAD repairs the link failure locally. The upstream node sends RREQ to the destination node which is located 7 hops away from it. Furthermore, the RREP is forwarded through the same number of hops after the destination responds by sending the RREP back to the upstream node. In the LOAD's local recovery, the probability of RREQ and RREP messages losses increases in high traffic and high number of hops environment. This increases the probability of local link repair failure, and hence increases the probability of packets losses.

On the other hand, when the 6RLR-ABC protocol locally repairs the broken link, it re-establishes the route to the destination through bypassing the failure node only. The upstream node sends a Local_RREQ message to the second next hop neighbour which is located nearby (either 1 or 2 hops) the upstream node.Hence, the Local_RREQ and Local_RREP messages do not pass through many hops between the upstream and second next hop nodes. Therefore, the probability of Local_RREQ and Local_RREP messages losses is reduced in this in low traffic and low number of hops environment. This reduces the probability of local link repair failure, and hence decreases the probability of packets losses. Table 4 and Figure 11 show that the 6RLR-ABC has 39% higher packet delivery ratio than LOAD protocol when traffic increases to 1200 packets.

Table 4. Packet Delivery Ratio versus Traffic

| Traffic | 6RLR-ABC Packet Delivery Ratio | LOAD Packet Delivery Ratio |
|---|---|---|
| 100 | 0.000000 | 0.000000 |
| 200 | 0.886000 | 0.773000 |
| 300 | 0.863500 | 0.739500 |
| 400 | 0.884000 | 0.744000 |
| 500 | 0.944000 | 0.786500 |
| 600 | 0.942200 | 0.773000 |
| 700 | 0.937167 | 0.757500 |
| 800 | 0.916000 | 0.728714 |
| 900 | 0.905250 | 0.707375 |
| 1000 | 0.898000 | 0.693556 |
| 1100 | 0.878000 | 0.654000 |
| 1200 | 0.862727 | 0.629273 |

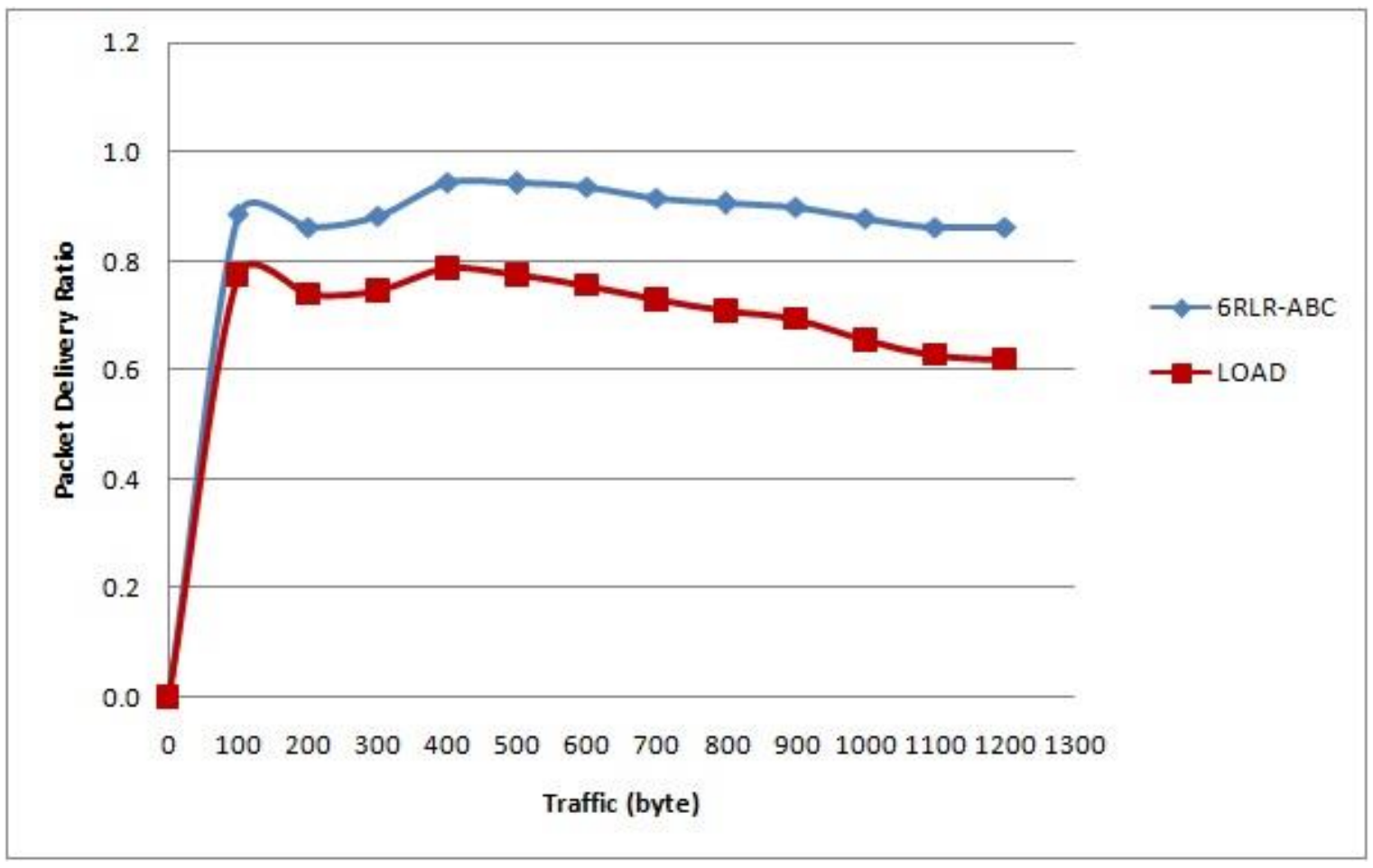

Figure 2. Packet Delivery Ratio versus Traffic

## 6.4 Average End-to-End Delay

The average end-to-end delay for local link repair mechanisms depends on two parameters. First, the number of hops between the nodes that are involved in the repair process (The initiators of both the RREQ and the RREP messages).Second, the traffic in the network. As explained earlier, when LOAD performs the local link repair to re-establish the route, the RREQ and RREP messages travel through many hops (7 hops) to the destination and back to the upstream node. The required time to broadcast the RREQ message and to receive back the RREP message is relatively long since the number of hops between the upstream node and the destination is high. Hence, the local link repair time of LOAD is long. This time contributes to the average end-to-end delay of the packets that need to queue in the upstream node during the local link repair process. In addition, when the traffic is high, LOAD's overhead expands the contention environment since the local repair area is enlarged toward the far destination node. This increases the average end-to-end delay due to the added contention delay.

In contrast, when 6RLR-ABC performs the local link repair to re-establish the route, the Local_RREQ and the Local_RREP messages do not travel for many hops to the second next hop neighbour and back to the upstream node (usually 1 or 2 hops).The required time to broadcast the Local_RREQ message and to receive back the Local_RREP message is relatively short since the number of hops between the upstream node and the second next hop neighbour is low. Hence, the local link repair time of 6RLR-ABC is short. Hence, the packets do not need to queue in the upstream node for a long time during the local link repair process. In addition, when the traffic is high, 6RLR-ABC's overhead does not expand the contention environment since the local repair area is limited within the local vicinity up to the second next hop neighbour only. Hence, the contention delaydoes not significantly contribute to the average end-to-end delay.

Table 5 and Figure 12 show the average end-to-end delay versus the number of transmitted packets for both 6RLR-ABC and LOAD protocols. The figure depicts that; 6RLR-ABC achieves lower average end-to-end delay compared to LOAD protocol. It achieves 20% lower average end-to-end delay compare to LOAD when the traffic increases to 1200 packets.

Table 5. Average End-to-End Delay versus Traffic

| Traffic | 6RLR-ABCAverage End-to-End Delay (s) | LOADAverage End-to-End Delay (s) |
|---|---|---|
| 100 | 0 | 0 |
| 200 | 6.336789 | 8.064062 |
| 300 | 6.569959 | 8.388141 |
| 400 | 5.978520 | 7.523975 |
| 500 | 5.845004 | 7.572276 |
| 600 | 6.040598 | 7.586053 |
| 700 | 5.324263 | 6.324263 |
| 800 | 5.714678 | 7.078315 |
| 900 | 5.636995 | 7.000631 |
| 1000 | 5.648260 | 6.814927 |
| 1100 | 6.057709 | 7.603164 |
| 1200 | 5.816706 | 7.271252 |

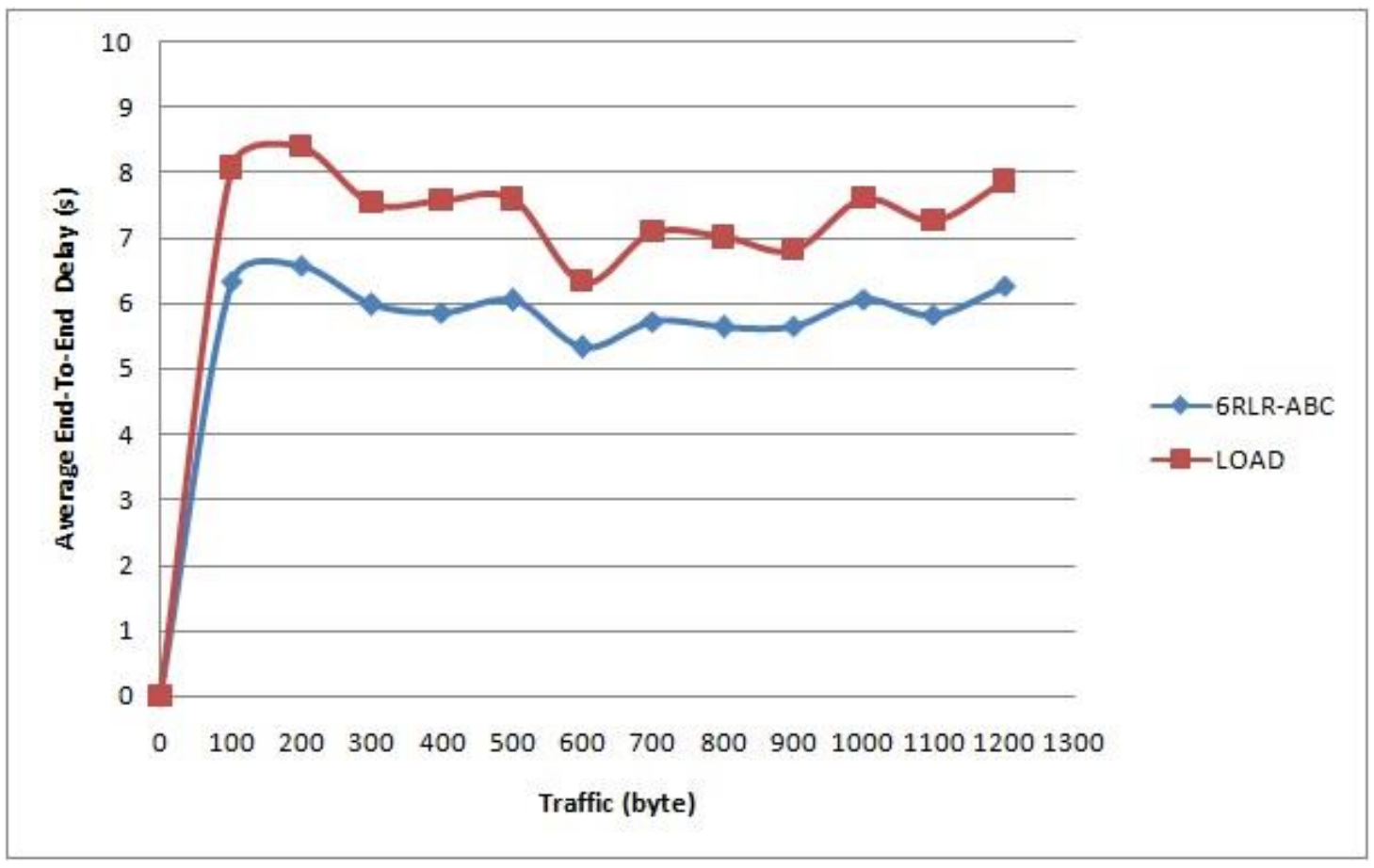

Figure 12. Average end-to-end delay versus traffic

## 7. CONCLUSIONS

6LoWPAN had been designed to enable the transmission of IPv6 packet over LoWPAN. Various 6LoWPAN routing protocols had been introduced to forward the packets between source and destination nodes in 6LoWPAN. However, 6LoWPAN routing protocols encounter few challenges including local link repair for the broken link. In this paper, 6LoWPAN Routing

Protocol with Local Repair Using Bio Inspired Artificial Bee Colony (6RLR-ABC) has been introduced.6RLR-ABC performs the local link repair through bypassing the failure node. After discovering the link failure, the upstream node repairs the link by re-establishing a new link to the second next hop neighbour. This new link bypasses the failure node. 6RLR-ABC protocol supports connection establishment between nodes in an energy-efficient manner while maintaining high packet delivery ratio and throughput and low average end-to-end delay. Simulation results have shown that the proposed 6RLR-ABC protocol outperforms LOAD protocol in terms of average energy consumption, throughput, packet delivery ratio, and average end-to-end delay. Thus, 6RLR-ABC has higher reliability than LOAD.

## ACKNOWLEDGEMENTS

This research was funded by the Deanship of Scientific Research at Princess Nourah bint Abdul rahman University through the Fast-track Research Funding Program.

## AUTHORS

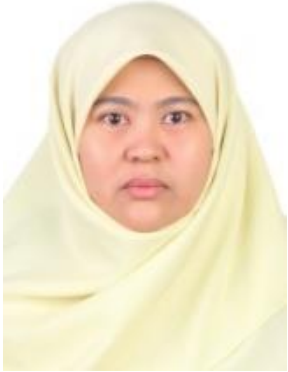

**Nurul Halimatul Asmak Ismail** received the degree in Computer Science from Universiti Sains Malaysia, in 2000 and master degree from Universiti Putra Malaysia, in 2009. She was working with Majlis Amanah Rakyat, Malaysia as a lecturer for Higher National Diploma in Computing an Edexcel program from United Kingdom. She involved with research group, program accreditation and syllabus construction. Years of experiences with Edexcel program encouraging her to finish her study in Ph.Dat 2015 in Computer Science specialization in Networking from Universiti Kebangsaan Malaysia, Bangi, Selangor, Malaysia. Recently, she joined Princess Nourah Bint Abdulrahman University, Kingdom of Saudi Arabia as assistant professor in department of Computer Science and Information Technology, College of Community. Her interests are in 6LoWPAN routing protocol, Internet of Things (IoT) and machine learning.

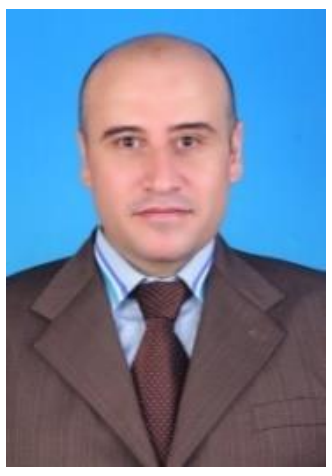

**Samer A. B. Awwad** received his B.Sc. in Engineering Technology with a major in Computer Engineering from Yarmouk University, Irbid, Jordan, in 2004. He worked for Technical and Vocational Training Corporation (Jeddah Military and Vocational Training Institute), Kingdom of Saudi Arabia, as a lecturer for 3 years. He received his M.Sc. in Communications and Network Engineering from Universiti Putra Malaysia, Serdang, Selangor, Malaysia, in 2010. He joined Nilai University as a lecturer from 2010 to 2013. He received his Ph.D. in Communications and Network Engineering from Universiti Putra Malaysia in 2016. He was a lecturer in the Department of Computer Engineering and Computer Science, Manipal International University, Malaysia between October 2015 and March 2019. He recently joined Fatimah for Information Technology Company as a consultant. His research interests include wireless ad hoc and sensor network specialized in mobility environments, 6LoWPAN, IEEE 802.15.4, IoT, compression and routing.

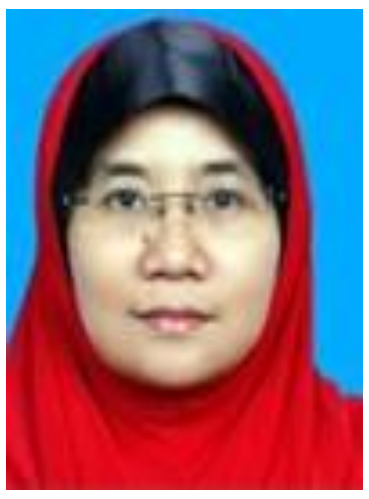

**Assoc Prof.Ts Dr .Hjh. Rosilah Hassan**, *SMIEEE* received her first degree from Hanyang University, Seoul, Republic of Korea in Electronic Engineering (1996). She works as an Engineer with Samsung Electronics Malaysia, Seremban before joining Universiti Kebangsaan Malaysia (UKM) in 1997. She obtains herMaster of Electrical Engineering (M.E.E) in Computer and Communication from UKM in 1999. In May 2008, she received her PhD in Mobile Communication from University of Strathclyde. Her research interest is in mobile communication, networking, 3G, and QoS. She is a senior lecturer at UKM for more than 20 years.